\documentstyle[11pt]{article}

\newcommand{\gsim}{{\mbox{\small{$\; {\mbox{\raisebox{0.48ex}{$>$}}}
        {\mbox{\hspace{-1.70ex}}}{\mbox{\raisebox{-0.48ex}{$\sim$}}}\;$}}}}

\newcommand{\nin}{{\mbox{$\;\in$
        {\mbox{\hspace{-1.70ex}}}$|\;$}}}

\begin{document}
\bibliographystyle{unsrt}

\newpage
\thispagestyle{empty}
{\mbox{ }}\\
{\vspace{-4.5 cm}}{\mbox{ }}\\
{\mbox{ }}{\hspace{10 cm}}UPR-0547-T\\
{\mbox{ }}{\hspace{10 cm}}December 1992\\
{\mbox{ }}\\ {\mbox{ }}\\ {\mbox{ }}\\

\begin{center}
{\LARGE{\bf Dark matter and  non-Newtonian gravity
from General Relativity on a
 stringy background }}\\
{\mbox{ }}\\
{\mbox{ }}\\
\renewcommand{\thefootnote}{\fnsymbol{footnote}}
Harald H.\ Soleng\footnote{On leave from
the University of Oslo, Norway}\footnote{Email address:
SOLENG@Steinhardt.hep.upenn.edu or
SOLENG@vuoep6.uio.no}\\
{\em Department of Physics\\
University of Pennsylvania\\
209 South 33rd Street\\
Philadelphia, Pennsylvania 19104-6396\\
 U.\ S.\ A.}
\end{center}
{\mbox{ }}\\
{\mbox{ }}\\

\begin{quote}
{\bf Abstract:}
An exact solution of Einstein's field equations for
a static spherically symmetric medium with a radially
boost invariant  energy-momentum tensor is presented.
In the limit of an equation of state corresponding to
 a distribution of radially directed strings
there is a $1/r$ correction to Newton's force law.
 At large distances and small accelerations this
law coincides with the phenomenological
force law invented by Milgrom in order to explain
the
flat rotation curves of galaxies without introducing dark matter.
The present model explaines why the
critical acceleration of Milgrom is of the same order of magnitude
as the Hubble parameter.
\end{quote}

\newpage
\addtocounter{footnote}{-\value{footnote}}

\section{Introduction}

Einstein's theory has had remarkable success in explaining
observed and inferred gravitational phenomena. There seems to
be only one serious problem --- the missing mass problem.
On large scales, the scales of galaxies and beyond, the
Einstein/Newton
dynamics
seems to imply that there is
much more mass than we can observe directly.
{}From observations of the rotational velocities of the
gaseous component of galaxies it is found that
the velocity approaches a constant at large distances (Sancisi \&
van~Albada 1987), and
from the relation
$v^2/r=g$, one finds that
the gravitational acceleration decreases as $1/r$ here.
 If luminuous matter were a good tracer
of mass and Newton's law were valid at these scales one should
find $g\sim 1/r^2$.

There are two explanations to this discrepancy. ($a$) Newtonian dynamics
is wrong at these scales, or ($b$) there is a lot of unseen ``dark matter"
in the galaxies.  Many authors have
advocated the first explanation and modifications of Newton's
gravitational dynamics have been proposed (Bekenstein \& Milgrom 1984;
Finzi 1963; Milgrom 1983a; Kuhn \&  Kruglyak 1987;
Liboff 1992;
Sanders 1990). Milgrom's theory (for reviews see: Milgrom 1987, 1989;
Milgrom \& Bekenstein, 1987)
has worked impressingly well
both for galaxies (Milgrom 1983b, 1984, 1986)
and galaxy systems (Milgrom 1983c),
and has recently
been found to be the best phenomenological description of
the systematics of the mass discrepancy in galaxies (Begeman, Broeils, \&
Sanders 1991; see however Gerhard \& Spergel 1992). Another argument
in favor of an effective $1/r$ correction to the force law at large
distances, is that such a term could stabilize a cold
stellar disk in a numerical galaxy model  (Tohline 1983).

In spite of the phenomenological success of non-Newtonian dynamics
the scientific community has been reluctant to
abolish Newton's theory of gravitation, partly because
Newton's theory is far more aesthetically attractive
than any of the modified theories, and since the
interactions appear to
be more fundamental than matter, one would
rather introduce new matter than new forces, but above all
it is objected that none of the modified theories have a viable
relativistic counterpart (Lindley 1992; Milgrom 1989).
Bekenstein's (1988a, b) phase coupling gravitation is consistent with
both extragalactic systematics and solar system tests, but at least
in the version with a sextic scalar potential (Bekenstein 1988a) it
has tachyonic propagation of scalar waves, and
it is unclear whether the model can be saved (Sanders 1990).
In general if a viable
 scalar-tensor theory that mimics Milgrom's exists, it
will be very complicated and contain many new
fundamental constants.
Therefore, the most widely accepted explanation is that
the Universe is filled with huge amonts of dark matter.

Recently it has been suggested that quantum gravity effects
may account for some of the missing mass (Goldman et al.\ 1992)
through a large scale variability of the effective $G$.
Here the idea of attributing dark matter to a very
specific and simple underlying principle is followed up, but
instead of a
variable $G$ from quantum gravity, we
propose a {\em stringy aether} as a specific form of dark matter.
In the case of spherical symmetry this energy-momentum tensor
re\-pro\-duces
the pheno\-meno\-logical force law of Milgrom (1983a)
within the
framework of classical General Relativity.
The
appearance of the Hubble constant in the modified
Newtonian force law is explained as a consequence of the
stringy nature of the dark matter.

\section{Radial boost invariance}

Newton's inverse square law follows from the weak field limit of
Einstein's theory
when one assumes that the energy-density of empty space is
exactly zero.
Einstein (1917) was the first to allow for  different large
scale dynamics when he introduced the cosmological constant, $\Lambda$,
in order to
get a long range repulsive force to balance the Newtonian
attraction and produce a static cosmological model.
Later the $\Lambda$-term has been understood as equivalent to the
energy-density of a maximally symmetric
vacuum (Gliner 1966). In fact, $\Lambda g_{\mu\nu}$
is the unique form of the energy-momentum
tensor if one assumes that the energy momentum tensor is
boost invariant in all directions.

In the empty space outside a star or a galaxy
one expects that isotropy, translational invariance, and
part of the boost invariance
of the vacuum is broken.
However, in order to retain as much as possible of the spirit
of the Strong Equivalence Principle\footnote{
The Strong Equivalence Principle states that the result of any local
test experiment, gravitational and nongravitational, in a freely falling
inertial frame does not depend on where and when it is performed
and not on the velocity of the inertial system.}
one would desire that boost invariance is preserved at least
in the radial direction, for only in this case would
the energy density be independent of the infall velocity of
the observer.
Then a freely falling observer would in principle
be unable to measure his radial velocity relative to the vacuum.
It is felt that this symmetry is the closest
one can get to the Strong Equivalence Principle
 in the case of
an anisotropic vacuum energy.
Accordingly, we will here take as our fundamental assumption that
 the medium outside a point mass has an effective
energy-momentum tensor which is invariant under radial boosts.
Thus, we will assume
$T^{t}_{\;\;t}=T^{r}_{\;\;r}$.
No assumptions will be made concerning the translational invariance
of the medium outside a point mass.
In the next section
the exact, static, spherically symmetric solution of Einstein's
field equations for radially boost invariant
energy-momentum tensors with angular components
proportional to the energy density is found.

\section{Exact solution}
Consider a static, spherically  symmetric space-time.
Up to coordinate transformations, we may write the metric as
\begin{equation}
ds^2=-e^{2\mu}dt^2+e^{2\lambda}dr^2+r^2d\theta^2
+r^2\sin^2{\theta }\, d\phi^2  \label{metrikk}
\end{equation}
where $\mu$ and $\lambda$ depends on the radial coordinate $r$ only.
With this metric  the Ricci tensor takes the form
\begin{eqnarray}
R^{t}_{\;\;t}&=&
\left[    -\mu''+\lambda'\mu'-\mu'^2-2\frac{\mu'}{r}
\right] e^{-2\lambda} \\
{\mbox{ }} & & {\mbox{ }}\nonumber\\
R^{r}_{\;\;r}&=&
\left[    -\mu''+\lambda'\mu'-\mu'^2+2\frac{\lambda'}{r}
\right] e^{-2\lambda}\\
{\mbox{ }} & & {\mbox{ }}\nonumber\\
R^{\Omega}_{\;\;\Omega}&=&
\left[-\frac{\mu'}{r}+\frac{\lambda'}{r}-\frac{1}{r^2}\right]e^{-2\lambda}
+\frac{1}{r^2}
\end{eqnarray}
where $\Omega$ stands for both $\theta$ and $\phi$.

With radial boost invariance and spherical symmetry,
the gravitational field is uniquely determined once
an ``equation of state'' specifies the relation between
$T^{t}_{\;\;t}$ and $T^{\Omega}_{\;\;\Omega}$.
Here we will assume that $T^{t}_{\;\;t}$ is proportional to
$T^{\Omega}_{\;\;\Omega}$.
Thus we get an
energy-momentum tensor of the form
\begin{equation}
T^{t}_{\;\;t}=T^{r}_{\;\;r}=-\alpha T^{\Omega}_{\;\;\Omega}\;\;\; ,
\label{Ansatz}
\end{equation}
where $\alpha$ is a finite dimension\-less constant.
With this {\em Ansatz} Einstein's field
equations\footnote{In this paper
geometrical units are used, i.e.\ $G=c=1$.}
\begin{equation}
G^{\mu}_{\;\;\nu}\equiv R^{\mu}_{\;\;\nu}-\frac{1}{2}g^{\mu}_{\;\;\nu}R
=8\pi  T^{\mu}_{\;\;\nu}\;\; ,
\end{equation}
imply
\begin{equation}
G^{t}_{\;\;t}=G^{r}_{\;\;r}\;\;\;{\mbox{ and }}\;\;\;
G^{t}_{\;\;t}=-\alpha G^{\Omega}_{\;\;\Omega}\;\; .
\end{equation}
{}From the first equation we find $\lambda=-\mu$.
Then the second equation becomes
\begin{equation}
\frac{1}{r^2}\left[( r e^{2\mu})'-1\right]=
-\frac{\alpha}{2r}\left[( r e^{2\mu})'-1\right]'\;\; .
\label{secondeq}
\end{equation}
The case $\alpha=0$ corresponds to the Schwarzschild solution.
For $\alpha\neq 0$, integration yields
\begin{equation}
( r e^{2\mu})'-1=-\epsilon\left(\frac{\ell}{r}\right)^{2/\alpha}
\label{secondint}
\end{equation}
where $\ell$ is a positive integration constant of dimension
length and $\epsilon=\pm 1$ is a sign factor which determines the sign of the
energy-density of the anisotropic vacuum ($\epsilon=1$ corresponds to
a positive energy density). Integrating once more, we find
\begin{equation}
e^{2\mu}=1-\frac{2M}{r}-\left\{  \begin{array}{lcl}
\epsilon\ell r^{-1}\ln{(\lambda r)} &
{\mbox{ for }} & \alpha=2\\
{\mbox{ }}&
{\mbox{ }}&
{\mbox{ }}\\
\epsilon\alpha(\alpha-2)^{-1}\ell^{2/\alpha}r^{-2/\alpha} &
{\mbox{ for }} & \alpha\neq 2
\end{array}
\right.  \label{solution}
\end{equation}
The following special cases are singled out: $\alpha=-1$ corresponds to the
Schwarzschild--de~Sitter solution, $\alpha=0$
corresponds to the
Schwarzschild solution, and  $\alpha=1$ corresponds to the
Reissner-Nordstr{\"o}m solution.

In the generic case,  $\alpha\nin\{0,2\}$,
the classical gravitational
acceleration is
\begin{equation}
g=\frac{M}{r^2}+\epsilon\frac{\ell^{-1}}{(\alpha-2)}
\left(\frac{\ell}{r}\right)^{1+2/\alpha}\;\; .
\label{graviacc}
\end{equation}

The energy-density, $\rho_{s}$, corresponding to
these solutions are found from
equations (\ref{secondeq}) and (\ref{secondint}). Hence, using that
$8\pi\rho_{s}=-8\pi T^{t}_{\;\;t}=-G^{t}_{\;\;t}$, one finds
\begin{equation}
8\pi \rho_{s}= \frac{\epsilon}{r^2}\left(\frac{\ell}{r}\right)^{2/\alpha}
\;\; .
\end{equation}

\section{String-like background and dark matter}

{}From equation (\ref{graviacc}) one gets
an effective  attractive
 $1/r$ correction to Newton's force if
$\epsilon\alpha\gg 1$.
In the present context a large $\alpha$ means that
$|T^{t}_{\;\;t}|=|T^{r}_{\;\;r}|\gg |T^{\Omega}_{\;\;\Omega}|$.
This may be understood  as the energy-momentum tensor of a cloud of
{\em radially directed strings at low but nonzero temperature}.

A straight string has vanishing gravitational mass, because
the gravitational effect of tension exactly cancels the
effect of its mass. Note that
if we let $\alpha\rightarrow \infty$ the correction
term in the metric coefficient of Equation
(\ref{solution}) becomes a constant, and thus
in the zero temperature limit the strings do not produce any
gravitational forces.
Assume that
the string has a mass
$m_{s}$. If this mass is evenly
distributed along the string which is streching across the whole universe,
the mass per length is $\mu = m_{s}H_{0}$.
Thus the energy gained by falling into a galaxy is
\begin{equation}
\Delta E = \frac{M\mu r}{r}=m_{s}MH_{0}\;\;\ .  \label{gravenergy}
\end{equation}
This energy is available to produce transverse motion of the string.
Hence, there will be a transverse pressure given by
\begin{equation}
(pV)^2=(m_{s}+\Delta E)^2-m_{s}^2\;\; . \label{press}
\end{equation}
The transverse pressure will
contribute with a pressure per energy density
$pV/m_{s}\approx(MH_{0})^{1/2}$.
This determines the dimensionless constant in the {\em Ansatz}
(\ref{Ansatz}) as follows
\begin{equation}
\alpha\approx \epsilon \left(MH_{0}\right)^{-1/2}\;\; .
\end{equation}
Hence, for positive $\rho_{s}$ ($\epsilon=1$), the angular pressure is
also positive. This agrees with the general lore that pressure contributes to
gravitational attraction.
The predicted  value of $|\alpha|$ is so large that
$(\ell/r)^{2/\alpha}\approx 1$
for all reasonable values of $\ell$. Hence,  with this information, Equation
(\ref{graviacc}) implies that
Newton's force law is changed to
\begin{equation}
g=\frac{M}{r^2}+k_{0}\frac{(MH_{0})^{1/2}}{r}  \label{forcelaw}
\end{equation}
where the constant $k_{0}\approx 1$.
The presence of the
Hubble parameter in the local force law signals that the
translational invariance of the background {\em aether}
is broken not only in spatial directions
but also in the time-direction. This is what one would
expect by including
strings of cosmic extension in the background {\em aether.}

Note that the force law of Equation (\ref{forcelaw})
agrees with the Tully-Fisher law (1977) which
relates the rotational velocity, $v$, to the luminosity, $L$,
by $v\propto L^{1/4}$ if the luminosity is proportional to the
Newtonian mass. This is a reasonable assumption
if the ratio of dark and luminous matter densities is a constant.
Also the mysterious coincidence that Milgrom's critical
acceleration is equal to the
Hubble parameter (Milgrom 1983a, 1989),  is explained as a result of having
dark matter of cosmic extension.

It could be objected that most observations seem to imply that the
mass density of the universe is smaller than the closure density, and that
the concept of strings
extending across the universe therefore is meaningless.
Note, however, that an isotropic  stringy background
has an energy-density of the form (Vilenkin 1985)
\begin{equation}
8\pi \rho_{s}=\frac{3w}{R^2}
\end{equation}
where $R$ is the cosmic scale factor and $w$ is a constant.
Accordingly, the first Friedmann equation takes the form
\begin{equation}
H^2= \frac{8\pi}{3}\rho+\frac{w-k}{R^{2}}\;\;.
\end{equation}
Hence, if one neglects the stringy
background the effective curvature is
$k_{{\mbox{{\footnotesize{eff}}}}}=k-w$ rather than $k$.
Thus a closed universe with a geometric $k\gsim 0$
as predicted by inflation
could be in agreement with the observed
$k_{{\mbox{{\footnotesize{eff}}}}}<0$ if the Universe
has a stringy background with $w>0$ (Vilenkin 1985).

A string dominated universe
requires that the intercommuting probability of the strings
 is very small, i.e.\ the
strings pass freely through one another. The strings
of a string dominated universe
could have a very small
mass per length $G\mu\sim 10^{-30}$, and even when passing through the
observer they would be difficult to detect
(Vilenkin 1984).

\section{Conclusion}
Up to now the missing mass problem has been resolved by assuming that
dark matter is present in whatever quantities and distributions that are
needed to explain away all mass descrepancies. The main problem with this
approach is that the dark matter hypothesis in this form
is too flexible to give
any unavoidable predictions (Milgrom 1989), and it is in principle not
testable before
one specifies the nature of the dark matter.
In contrast,
the approach of Milgrom is testable, and it
gives specific predictions which are in good agreement with
observations. The main problem has been that the modified dynamics
has no viable relativistic counterpart.

Here it has been shown that Einstein's General Relativity coupled to
a {\em stringy aether} reproduces the force law of Milgrom (1983a).
In a closed universe this model explaines the Machian character of
Milgrom's acceleration, $a_{0}\approx H_{0}$,
as a consequence of a {\em stringy aether} extending over
the whole universe.
The dark matter model presented here
 is a phenomenological model,
but it can no longer be objected that
the $1/r$ modification of Newton's force law is ``an orphan in the classical
world'' (Lindley 1992). Instead it follows from General Relativity with
a relativistic energy-momentum tensor corresponding to
a {\em stringy aether}.

Despite the success of Milgrom's force law and the fact that the present
model reproduces it, there are many reasons
why it is premature to identify a {\em stringy aether} as
{\em the solution} to the missing mass problem. First, there is no
field theoretic realization of this particular model. Second,
the proposal has only been studied in a static, spherically symmetric
model. In contrast, the real universe is non-static and it contains
many galaxies so both of the symmetry assumptions are broken.
It is not clear how deviations from spherical
symmetry will affect the solution,
and especially how  strings passing through more than one galaxy will affect
the model. Finally one expects realistic string models to predict
that strings will intercommute and produce closed loops.

However, from a general relativistic perspective it is very interesting
that such a simplistic model can reproduce the non-Newtonian force law.
The generally accepted solution --- dark matter --- need therefore not be
a very complicated system of epicycles as have been argued by
the proponents of non-Einsteinian gravitation (e.g.\ Sanders 1990).
\subsection*{Acknowledgements}
I wish to thank Paul J.\ Steinhardt for stimulating my interest in
the dark matter problem and for valuable comments.
This research was supported in part by
the Fridtjof Nansen Foundation Grant No.\ 152/92, by
Lise and Arnfinn Heje's Foundation, Ref.\ No.\ 0F0377/1992, by
the Norwegian Research Council for Science and the Humanities
(NAVF), Grant No.\ 420.92/022,
and by the U.S.\ Department of Energy under Contract
No.\ DOE-EY-76-C-02-3071.

\newpage
\section*{References}
Begeman, K.\ G., Broeils, A.\ H., \& Sanders, R.\ H. 1991,
Mon.\ Not.\ Roy.\ Astron.\ Soc., {\bf 249}, 523\\
{\mbox{ }}\\
Bekenstein, J. 1988a, in {\em Proc. of 2nd Canadian Conf.\ on
General Relativity and Relativistic Astrophysics},
ed.\ A.\ Coley, C.\  Dyer, \& T.\ Tupper
(Singapore: World Scientific Publishing Co.\ Pte.\ Ltd.), p68\\
{\mbox{ }}\\
Bekenstein, J. 1988b, Phys.\ Lett.\ B, {\bf 202}, 497\\
{\mbox{ }}\\
Bekenstein, J., \& Milgrom, M. 1984,
Astrophys.\ J., {\bf 286}, 7\\
{\mbox{ }}\\
Einstein, A. 1917, Preuss.\ Akad.\ Wiss.\ Berlin Sitzungsber., 142\\
{\mbox{ }}\\
Finzi, A. 1963, Mon.\ Not.\ Roy.\ Astr.\ Soc., {\bf 127}, 21\\
{\mbox{ }}\\
Gerhard, O.\ E., \& Spergel, D.\ N. 1992,
Astrophys.\ J., {\bf 397}, 38\\
{\mbox{ }}\\
Gliner, {\'E}.\ B. 1966, Sov.\ Phys.\ JETP, {\bf 22}, 378\\
{\mbox{ }}\\
Goldman, T., P{\'e}rez-Mercader, J., Cooper, F., \& Nieto, M.\ M.
1992, Phys.\ Lett.\ B {\bf 281}, 219\\
{\mbox{ }}\\
Kuhn, J.\ R., \& Kruglyak, L. 1987,
Astrophys.\ J., {\bf 313}, 1\\
{\mbox{ }}\\
Liboff, R.\ L. 1992,
Astrophys.\ J.\ Lett., {\bf 397}, L71\\
{\mbox{ }}\\
Lindley, D. 1992, Nature {\bf 359}, 583\\
{\mbox{ }}\\
Milgrom, M. 1983a, Astrophys.\ J., {\bf 270}, 365\\
{\mbox{ }}\\
Milgrom, M. 1983b, Astrophys.\ J., {\bf 270}, 371\\
{\mbox{ }}\\
Milgrom, M. 1983c, Astrophys.\ J., {\bf 270}, 384\\
{\mbox{ }}\\
Milgrom, M. 1984, Astrophys.\ J., {\bf 287}, 571\\
{\mbox{ }}\\
Milgrom, M. 1986, Astrophys.\ J., {\bf 302}, 617\\
{\mbox{ }}\\
Milgrom, M. 1987, in {\em Dark Matter in the Uni\-verse:
proceedings of the 4th Jerusalem Winter School for Theoretical
Physics}, ed.\ J.\ N.\ Bahcall,  T.\ Piran,  \& S.\ Weinberg
(Singapore: World Scientific Publishing Co.\ Pte.\ Ltd.), p231\\
{\mbox{ }}\\
Milgrom, M. 1989, Comm.\ Astrophys., {\bf 13}, 215\\
{\mbox{ }}\\
Milgrom, M., \& Bekenstein, J. 1987, in
{\em Dark Matter in the Universe} 117th IAU symposium,
ed.\ H.\ Kormendy \&
G.\ R.\ Knapp
(Dordrecht: D.\ Reidel Publishing Co.), p319\\
{\mbox{ }}\\
Sancisi, R., \& van~Albada, T.\ S. 1987, in
{\em Dark Matter in the Universe} 117th IAU symposium,
ed.\ H.\ Kormendy \&
G.\ R.\ Knapp
(Dordrecht: D.\ Reidel Publishing Co.), p67\\
{\mbox{ }}\\
Sanders, R.\ H. 1990, Astron.\ \&  Astrophys.\ Rev., {\bf 2}, 1\\
{\mbox{ }}\\
Tohline, J.\ E. 1983, in {\em Internal Kinematics and Dynamics of
Galaxies} 100th IAU symposium, ed.\ A.\ Athanassoula (Dordrecht: D.\
Reidel Publishing Co.), p205\\
{\mbox{ }}\\
Tully, R.\ B., \& Fisher, J.\ R. 1977,
Astron.\ \& Astrophys., {\bf 54}, 661\\
{\mbox{ }}\\
Vilenkin, A. 1984, Phys.\ Rev.\ Lett., {\bf 53}, 1016\\
{\mbox{ }}\\
Vilenkin, A. 1985, Phys.\ Rep., {\bf 121}, 263\\
{\mbox{ }}

\end{document}